\documentclass{kristall}               

\usepackage{epsf}
\usepackage{bm}

\begin{document}
\maketitle              

\begin{abstract}        
The high pressure behaviour of 3R-NbS$_2$ has been investigated by
angle-dispersive X-ray powder diffraction using diamond anvil cells up
to 14~GPa. The compression behaviour of the
structure is highly anisotropic. The compressibility perpendicular to
the layers is 2.5 times higher than within the layers. A fit of a
3$^{rd}$-order Birch-Murnaghan equation of state gave a volume at zero
pressure V$_0$=174(1)~\AA$^3$ and a  bulk modulus
b$_0$=57(1) GPa, with a pressure derivative b'=8.6(5).  
\end{abstract}          


\section*{Introduction}

Transition metal dichalcogenides and their intercalation compounds
attracted significant attention due to their superconductivity
\cite{Jerome} and their potential use in solid state batteries
\cite{Whittingham} and catalytic processes in 
petrochemical industry \cite{Chianelli}. NbS$_2$ crystallises in a
quasi two dimensional structure. Layers of trigonal edge-sharing
NbS$_6$ prisms are stacked onto each other along the [001]
direction. The intra-layer bonding is predominantly covalent, whereas
the layers are connected by van-der-Waals forces. Different stacking
sequences of the NbS$_2$ layers lead to the formation of two polytypes,
the hexagonal 2H-NbS$_2$ \cite{Jellinek} with two NbS$_2$ layers, and
the rhombohedral 3R-NbS$_2$ \cite{Morosin74,Powell81} with three
layers per unit cell, according to the modified Gard notation
\cite{IUCR}. The 3R polymorph crystallises in the space
group $R3m$ (no. 160) with the unit cell parameters of $a$=3.3285(4) and
$c$=17.910(4) \AA. All atoms are located on the Wyckoff position
3$a$ (0,0,z) with z=0 for niobium and z=0.2464(1) and 0.4201(2) for
the sulphur atoms, respectively \cite{Morosin74,Powell81}.
The layer structure causes a strong anisotropy in the physical
properties e.g. resistivity and compressibility \cite{Jerome,Jones72}.
At ambient conditions NbS$_2$ is a metal \cite{Krasowski}.

The effect of pressure on the critical temperature of
superconductivity, as well as the compressibility of 3R-NbS$_2$ has
been investigated up to 1 GPa  \cite{Jones72}. The authors found the linear
compressibilities k$\|$a=0.0016(4) GPa$^{-1}$ and k$\|$c=0.011(4)
GPa$^{-1}$ from their X-ray diffraction measurement using a
piston-cylinder apparatus and a slight pressure dependence of the
superconducting transition temperature T$_c$. However, the extension
of the pressure range beyond 1 GPa is of interest, because 
pressure leads to changes in the electronic bands, which may be associated
with a structural phase transition \cite{Akbarzadeh}.

Here we present the results of a high-pressure X-ray powder
diffraction study on 3R-NbS$_2$ up to 14 GPa.

\section*{Experimental}

The sample of 3R-NbS$_2$ was prepared by sintering a stoichiometric
mixture of the elements for 10 days at 1125~K. The composition of
the sample was verified by electron micro-probe analysis
using a Cameca ``Camebax Microbeam''. The polytype of the sample was
determined by conventional powder diffraction.
The high pressure data were collected with a MAR~2000 image
plate diffractometer with a wavelength of
0.5608~\AA \ (Ag K$_{\alpha}$). A silicon (111)
monochromator was applied in the incident beam for the suppression of the background and
$\lambda$/2 reflections from the gasket material. Pressure was applied through a
Merrill-Bassett type diamond anvil cell \cite{Merrill} with an Inconel
gasket. As the pressure transmitting medium a 16:3:1 mixture of
methanol:ethanol:water was used \cite{Fujishiro82}. The ruby fluorescence
method using the pressure scale of Piermarini {\it et
al.} \cite{Piermarini75} was used to determine the pressure. 
The sample to detector distance was determined by measuring the sample
in the uncompressed cell. The geometry correction for the radial
integration of the two-dimensional data and the transformation into
standard one-dimensional powder patterns were performed using 
{\sc Fit2d} \cite{Hammersley96}. The single reflections were fitted
with Gaussian peak shape function
and a 3$^{rd}$-order polynom for the background with the program {\sc
  MFit} \cite{Hammersley89}. The unit cell parameters were calculated
using the refinement program {\sc Refcel} \cite{Cockcroft}.

\section*{Results and Discussion}

\begin{figure}[h!]
\begin{minipage}[b]{1.0\linewidth}
\epsfxsize=\linewidth
\noindent
\epsfbox{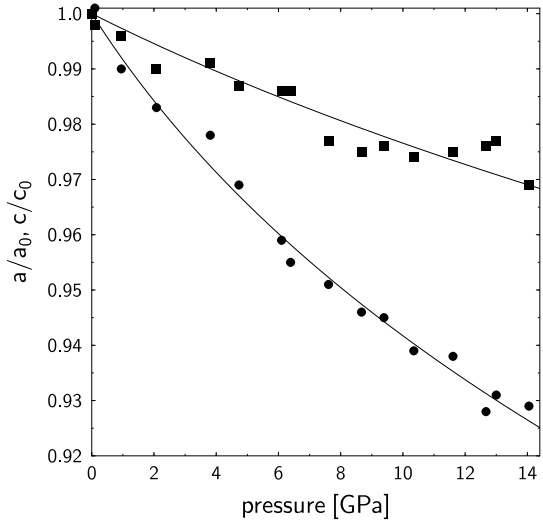}
\end{minipage}\hfill
\begin{minipage}[b]{.95\linewidth}
\noindent
\caption[]{\label{fig-ac} Variation of the normalised cell parameters
  for 3R-NbS$_2$ with pressure. The squares correspond to a/a$_0$ and
  the circles represent c/c$_0$. The experimental errors bars
  correspond to the size of the symbols. The lines are guides for the eyes.}
\end{minipage}
\end{figure}

\begin{figure}[h!]
\begin{minipage}[b]{.95\linewidth}
\epsfxsize=\linewidth
\noindent
\epsfbox{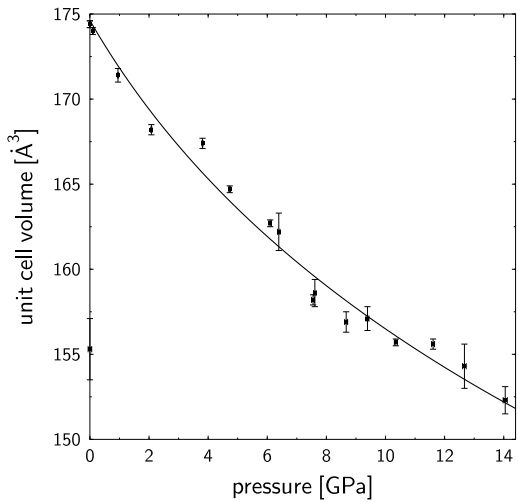}
\vspace*{0pt}
\end{minipage}\hfill
\begin{minipage}[b]{1.0\linewidth}
\noindent
\caption[]{\label{fig-pv} Pressure dependence of the unit cell volume of
  3R-NbS$_2$. The line is a fit of a 3$^{rd}$-order Birch-Murnaghan
  equation-of-state to the data points.}
\vspace*{0pt}
\end{minipage}
\end{figure}

Diffraction patterns were collected up to 14 GPa.
The pressure dependence of  the normalised
lattice parameters a/a$_0$ and c/c$_0$ and of the unit cell volume are given in figures \ref{fig-ac}
and \ref{fig-pv}, respectively.

A 3$^{rd}$-order Birch-Murnaghan equation-of-state was used to determine
the unit cell volume at pressure of 0~GPa, $V_0$, the bulk modulus
$b_0$ and its pressure dependence $b' = \partial b_0 / \partial p$
\cite{Birch78}. Values of V$_0$=174(1)~\AA$^3$, b$_0$=57(1) GPa
 and b'=8.6(5) were obtained. V$_0$ agrees well with the data given by
Morosin \cite{Morosin74} and Powell and Jacobson \cite{Powell81} for the
unit cell volume at ambient conditions. 
The linear compressibility $k$ was derived by fitting linear
functions to the unit-cell parameters $a$ and $c$ at pressures up to
1.5 GPa. The resulting values are given in table \ref{tab1}.
The compression behaviour is highly anisotropic with compressibility
parallel to the $c$ axis being 2.5 times higher than parallel the $a$ axis.  
The anisotropy of the compression is in the order of magnitude for
related transition-metal dichalcogenides e.g. TiS$_2$ \cite{Allan},
2H-MoS$_2$ \cite{Webb,Flack} and 2H-NbSe$_2$ \cite{Jones72,Flack} but less
pronounced than in SnS$_2$ \cite{Knorr} (see table \ref{tab1}). 

The data for the linear compressibilities for 3R-NbS$_2$ given by
Jones et al. \cite{Jones72} are in good agreement with our values for
the $c$ direction. However, the k$\|$a values differ by a factor of
two. The compressibility data for 2H-NbSe$_2$ given by Flack
\cite{Flack} and Jones et al. \cite{Jones72} are likewise
significantly different. Hence, one might attribute the differences in
the linear compressibilities along the $a$ axis to experimental
uncertainties in the work of Jones et al. \cite{Jones72}.

\begin{table}[ht!]
\caption[]{\label{tab1}Values for the linear compressibility
  parallel to the $a$ and $c$ axis for several metal dichalcogenides.}

\def\Strut{\large\strut}
\small
\begin{tabular*}{\linewidth}{@{\extracolsep{\fill}} @{}l  clc @{}}
\hline
 &k$\|$a [GPa$^{-1}$] & k$\|$c [GPa$^{-1}$]  & Reference \\
\hline
3R-NbS$_2$  & 0.0039(3) & 0.010(4)  & this work\\
3R-NbS$_2$  & 0.0016(4) & 0.011(4)  & \cite{Jones72}\\
1T-TiS$_2$  & 0.0043(3) & 0.017(2)  & \cite{Allan}\\
1T-SnS$_2$  & 0.0022(4) & 0.025(5)  & \cite{Knorr}\\
2H-MoS$_2$  & 0.0034(1) & 0.0164(3) & \cite{Webb}\\
2H-MoS$_2$  & 0.0033(1) & 0.0170(2) & \cite{Flack}\\
2H-NbSe$_2$ & 0.0041(4) & 0.0162(5) & \cite{Jones72}\\
2H-NbSe$_2$ & 0.0015(3) & 0.011(1)  & \cite{Flack}\\

\hline
\end{tabular*}

\end{table}

The strong anisotropy in the compression can be understood by 
considering of the different types of bonding in the structure.
The lowest compressibility is found parallel to the $a$ axis,
where the structure consists of layers of covalent
bonded trigonal NbS$_6$ prisms, whereas the layers are connected by
van-der-Waals forces along the $c$ direction. The compressibility
parallel to the stacking direction $c$ is 2.5 times stronger. 
This can be attributed to the ease with which the distance
between the van-der-Waals bonded layers can be reduced compared to distorting the covalent
bonded trigonal prisms which would require considerably more energy.
In the pressure range investigated, the data reveal that there is
no phase transition or discontinuity in the compression of 3R-NbS$_2$.

Unfortunately, the quality of the data was not sufficient for
a structure refinement. However, considering closely related
structures e.g. TiS$_2$ \cite{Allan} and SnS$_2$ \cite{Knorr} the
following compression mechanism is proposed: The reduction of the
van-der-Waals gap is the main compression mechanism, manifested in the
strong
change of the interlayer S-S distances. The lengths of covalent metal-sulphur
bonds in the layers should remain almost constant. Therefore, the compression in
the $a$ direction should result in slight changes in the metal-sulphur
angles in the trigonal prisms, resulting in an expansion of the 
layer-thickness under pressure. 
This mechanism is additionally supported by force field calculations on
2H-MoS$_2$ \cite{Webb}.

\begin{acknowledgment}
This research was performed within the framework of the Kieler
Forschergruppe on ``Wachstum und Grenzfl\"acheneigenschaften von
Sulfid- und Selenid-Schicht-Strukturen'' which is funded by the German
Science Foundation DFG (FOR 353/1-2 and De 412/21-1). We thank
Mrs. B. Mader for performing the microprobe analysis.
\end{acknowledgment}



\newpage

\begin{thebibliography}{99}
\bibitem{Jerome} J\'{e}rome, D.; Berthier, C.; Molini\'{e}, P.; Rouxel,
  J.: Electronic properties of transition metal
 dichalcogenides: Connection between structural instabilities and
  superconductivity. J. Phys. \textbf{4} (1976) 125-135
\bibitem{Whittingham} Whittingham, M.S.: Intercalation Chemistry and
  Energy Storage. J. Solid State Chem. {\bf 29} (1979) 303-310
\bibitem{Chianelli} Chianelli, R.R.; Daage, M.; Ledoux, M,J.:
  Fundamental Studies of Transition Metal Sulfide Catalytic Materials,
  Advances in Catalysis \textbf{40} (1994) 177-232
\bibitem{Jellinek} Jellinek, F. ;Brauer, G.; M\"uller, H.:
 Molybdenum and niobium sulphides, Nature {\bf 185} (1960) 376-377 
\bibitem{Morosin74} Morosin, B.: Structure refinement on NbS$_2$. Acta
  Cryst. B~\textbf{30} (1974) 551-552
\bibitem{Powell81} Powell, D.R.; Jacobson, R.A.: The crystal structure
  of 3R-Nb$_{1.06}$S$_2$. J. Solid State Chem. \textbf{37} (1981) 140-143
\bibitem{IUCR} Guinier, A.; Bokij, G.B.; Boll-Dornberger, K.; Cowley,
  M.; \v{D}urovi\v{c}, S.; Jagodzinski, H.; Krishna, P.; De Wolff,
  P.M.; Zvyagin, B.B.; Cox, D.E.; Goodman, P.; Hahn, Th.; Kuchitsu,
  K.; Abrahams, S.C.: Nomenclature of polytype structures. Acta Cryst.
A~\textbf{40} (1984) 399-404
\bibitem{Krasowski} Krasowski, R.V.: Band structure of MoS$_2$ and
  NbS$_2$. Phys. Rev. Lett. \textbf{37} (1973) 1175-1178
\bibitem{Jones72} Jones, R.E.; Shanks, H.R.; Finnemore, D.K.: Pressure
  effect on superconducting NbSe$_2$ and NbS$_2$. Phys. Rev. B
  \textbf{6} (1972) 835-838 
\bibitem{Akbarzadeh} Akbarzadeh, A.; Clark, S.J.; Ackland, G.J.:
A theoretical study of selenium I under high pressure. J. Phys.:
  Condens. Matter \textbf{5} (1993) 8065-8074
\bibitem{Merrill} Merrill, L.; Bassett, W.: Miniature diamond anvil
  pressure cell for single crystal x-ray diffraction studies.
Rev. Sci. Instrum.  \textbf{45} (1974) 290-294 
\bibitem{Fujishiro82} Fujishiro, I.; Piermarini, G.J.; Block, S.;
  Munro, R.G.: High Pressure in Research and Industry, Proceedings of
  the 8th AIRPT Conference Uppsala, edited by Backman, C.M.;
  Johannisson, T.; Tegner, L.; (1982) 608 
\bibitem{Piermarini75} Piermarini, G.J.;Block, S.; Barnett, J.D.; 
  Forman, R.A.: Calibration of the pressure dependence of the $R_1$
  ruby fluorescence line to 195 kbar. J. Appl. Phys. \textbf{47}
  (1975) 2774-2780
\bibitem{Hammersley96}  Hammersley, A.P.; Svensson, S.O.; Hanfland, M.;
  Fitch, A.N.; H\"ausermann, D.: Two-dimensional detector software:
  from real detector to idealised image or two-theta scan. High
  Pressure Research \textbf{14} (1996) 235-248
\bibitem{Hammersley89} Hammersley, A.P.; Riekel, C.: MFIT: Multiple
  spectra fitting program. Synchrotron Radiation News \textbf{2}
  (1989) 24-16 
\bibitem{Cockcroft} Cockcroft, J.: REFCEL Version 3.03, Collaborative
  computational project number 14 (CCP14) for single  crystal
  and powder diffraction. {\it www.ccp14.ac.uk}
\bibitem{Birch78} Birch, F.: Finite strain isotherm and velocities for
  single-crystal and polycrystalline NaCl at high pressure and 300
  K. J. Geophys. Res. \textbf{83} (1978) 1257-1268
\bibitem{Allan} Allan, D.R.; Kelsey, A.A.; Clark, S.J.; Angel, R.J.;
  Ackland, G.J.: High-pressure semiconductor-semimetal transition in
  TiS$_2$. Phys. Rev. B \textbf{57} (1998) 5106-5110
\bibitem{Knorr} Knorr, K.; Ehm, L.; Hytha, M.; Winkler, B.; Depmeier,
  W.: The High Pressure Behaviour of SnS$_2$: X-ray Powder Diffraction
  and Quantum Mechanical Calculations up to 10
  GPa. Phys. Stat. Sol. (b) \textbf{223} (2000) 435-440 
\bibitem{Webb} Webb, A.W.; Feldman, L.J.; Skelton, E.F.; Towle, L.C.;
  Liu, C.Y.; Spain, I.L.: High Pressure Investigations of MoS$_2$.
J. Phys. Chem. Solids \textbf{37} (1976) 329-335
\bibitem{Flack} Flack, H.D.: Compressibilities of some layer dichalcogenides
J. Appl. Cryst. \textbf{5} (1972) 137-138

\end{thebibliography}
\end{document}